\documentclass[prl,aps,showpacs,10pt]{revtex4}%
\usepackage{amsfonts}
\usepackage{amsmath}
\usepackage{amssymb}
\usepackage{graphics}
\begin{document}
\title{Excitation Induced Dephasing in Semiconductor Quantum Dots}
\author{H. C. Schneider}
\affiliation{Physics Department, Kaiserslautern University of Technology, P. O. Box 3049,
67653 Kaiserslautern, Germany}
\author{W. W. Chow}
\affiliation{Semiconductor Materials and Device Science Department, Sandia National
Laboratories, Albuquerque, NM 87185-0601}
\author{S. W. Koch}
\affiliation{Physics Department, Philipps University, Renthof 5, 35037 Marburg, Germany}

\begin{abstract}
A quantum kinetic theory is used to compute excitation induced
dephasing in semiconductor quantum dots due to the Coulomb interaction
with a continuum of states, such as a quantum well or a wetting
layer. It is shown that a frequency dependent broadening together with
nonlinear resonance shifts are needed for a microscopic explanation of
the excitation induced dephasing in such a system, and that excitation
induced dephasing for a quantum-dot excitonic resonance is different
from quantum-well and bulk excitons.

\end{abstract}
\maketitle

Semiconductor quantum dots (QDs) are widely studied, e.g. for quantum
optical applications,~\cite{michler:nature-quant-correlation,bayer:prl01:spont-em} quantum information processing,~\cite{steel:all-optical-gate} or semiconductor laser systems.~\cite{Sugawara-book, Bimberg-book} In all these optical
investigations, the excitation induced decoherence of the material
polarization driven by optical fields plays a decisive role. The
dephasing of the optical polarization limits all externally
controllable coherent and quantum entanglement processes, and
determines the homogeneous broadening of quantum-dot laser active
media.~\cite{uskov-jauho:dotdephasing-phonon,prb02:anomalous-dispersion,sugawara:homogeneous_broadening}
Measurements of polarization dephasing in self-organized QDs have
recently been reported,~\cite{borri:dotdephasing-prl} and excitation
induced dephasing processes have been observed in interface QDs, where
they were found to be important even at low temperatures and
densities.~\cite{guenther:quantum-dot-response}

Currently missing is a consistent microscopic theory for the
excitation induced dephasing of polarizations created between in QD
states due to Coulomb interaction with carriers in continuum
states. The traditional analysis of scattering and dephasing
processes in condensed matter systems with continuous energy spectra
invokes a Markovian approximation, in which the full time-dependence
of the dephasing process is neglected and replaced by an instantaneous
scattering event. The Markov approximation thus discards the memory
time of a scattering process, which corresponds to a
quantum-mechanical energy-time uncertainty, and describes dephasing by
equations of Boltzmann type using \textit{energy conserving
transitions} between quasi-particle states. The
continuous energy spectra of carriers in bands ensure that through the
energy-momentum relation in two or three dimensions a large phase
space contributes to the scattering/dephasing processes. Moreover,
scattering/dephasing processes for a continuum of states do not single
out one energy because the relevant physical quantities are
obtained by an average over a range of energies.

For the dephasing of the optical polarization in QD structures such a
treatment needs to be investigated because a discrete QD state enters
the energy conservation condition, which reduces the available phase
space by restricting the allowed scattering
possibilities.~\cite{toda:carrier-relaxation,urayama:prl01:phonon-bottleneck}
Disregarding inhomogeneous broadening, the dephasing affects only the
very narrow QD resonance, whose value is \textit{excitation
dependent}. Thus one has to take the frequency dependence of the
dephasing and the excitation-induced renormalization of the QD
resonances seriously. This is done in this paper using a quantum
kinetic analysis that avoids the Markovian approximation and thus
includes the memory of the interaction with carriers in scattering
states leading to polarization
dephasing.~\cite{banyai:exciton-lophonon} Another interesting aspect
is that the arguments made above against employing the Markov
approximation for a narrow resonance also pertains to excitons, i.e.,
Coulomb bound states. The non-Markovian calculation treats both types
of bound states (QD resonances and excitons) on an equal footing and
the different behavior of the two types of bound states can be
investigated.

In this letter we present a microscopic non-Markovian calculation of the
excitation induced polarization dephasing for a model representing shallow
epitaxially grown or interface QDs. The heterostructure is treated as a system
of QDs embedded in a quantum well (QW) and  the optical polarization is
computed including the Coulomb interaction between carriers in the dot states
and carriers populating the continuum of scattering states in the QW. The
Coulomb interaction causes excitation dependent
modifications of QD optical response to external fields: First, the
non-Markovian treatment of the Coulomb interaction between QDs and surrounding
QW leads to an energy dependence of the polarization dephasing. Second,
excitation-induced shifts of the QD resonance energies are obtained, in
agreement with experimental results. It is shown that the combination of
resonance-shifts and energy dependent polarization dephasing leads to an
excitation dependence of the QD optical response that is qualitatively
different from the QW and bulk systems. 

Our analysis starts from the standard many-particle Hamiltonian
including Coulomb interacting electrons and holes with dipole coupling
to a classical electromagnetic field.~\cite{kochbuch} Since we are
modelling QD and QW states interacting via the Coulomb
potential,~\cite{prb:many-body-qds} the relevant Coulomb interaction
energy matrix element
$V_{rsnm}=\sum_{\vec{q}\neq0}V_{q}I_{nm}(q)^{\ast }\ I_{rs}(q)$
contains the Fourier-transformed QW Coulomb potential $V_{q}$ and
overlap integrals $I_{rs}(q)=\int d^{2}r\phi_{r}^{\ast}\left( r\right)
e^{-i\vec{q}\cdot\vec{r}}\phi_{s}\left( r\right) $. Here
$\phi_{n}\left( r\right) $ is a carrier envelope function in the plane
of the QW, which can be localized or delocalized, describing a dot or
wetting layer state, respectively. The integral is extended over the
QW plane. We neglect the interaction with phonons because this
interaction mechanism gives an excitation-independent background
dephasing that can be separated from the excitation-induced
contribution.~\cite{borri:dotdephasing-prl}

We apply the nonequilibrium Green's functions
technique~\cite{schaefer-book,haug-jauho} to compute the linear
optical response using the free generalized Kadanoff-Baym ansatz
according to the general approach presented in
Ref.~\cite{binder-koch,khitrova:rmp99}. In order to illustrate the
physical effects due to the interaction between carriers in QD and QW
states, we treat the case of one localized state (denoted by
\textquotedblleft 0\textquotedblright\ for the QD ground state) that
is coupled to delocalized scattering states (denoted by $k$ for the
wetting layer or QW states) for electrons and holes. Typically, in
realistic systems different QDs are sufficiently far apart that there
is no coupling between dots at different positions. We furthermore
assume identical single-particle wave functions for electrons and
holes to avoid the distortion of the continuum states due to charged
dots. Then the dynamical variables are the polarizations of the QD,
$p_{0}=\langle b_{0}a_{0}\rangle$, and the QW, $p_{k}=\langle
b_{k}a_{k}\rangle$. For the evaluation of the probe absorption
spectra, we retain all terms linear in the induced probe polarization
to obtain dynamical equations for $p_{0}$ and $p_{k}$, in which the
dephasing is described by memory integrals over interaction processes
with the continuum. We do not employ a Markov approximation and
transform the dynamical equations to frequency space. The resulting
Fourier-transformed polarization equations with energy dependent
scattering rates contain the full memory of the polarization dynamics
in the time domain.

In detail, the electron-hole polarization connecting the electron and hole
ground states (\textquotedblleft0\textquotedblright)\ is determined by
\begin{equation}
(\hbar\omega-\epsilon_{0,\text{HF}}^{e}-\epsilon_{0,\text{HF}}^{h})p_{0}%
+\hbar\Omega_{00,\text{HF}}(1-n_{0}^{e}-n_{0}^{h})=iS_{0}(\omega)\label{p-om}%
\end{equation}
with the scattering contribution
\begin{equation}
S_{0}(\omega)=-\Lambda_{00}p_{0}(\omega)+\sum_{\vec{k}^{\prime}}%
\Lambda_{0,k^{\prime}}p_{k^{\prime}}(\omega).\label{S-om}%
\end{equation}
In Eq.~(\ref{p-om}), the Rabi energy $\hbar\Omega_{00,\text{HF}}$ and the QD
single-particle energies $\epsilon_{0,\text{HF}}^{\alpha}$ ($\alpha=$ e,h) are
renormalized energies containing the many-body Hartree-Fock contributions, and
the bare QD energies $\epsilon_{0}^{\alpha}$ contain the QD electron-hole
binding energy. The carrier distribution functions are denoted by
$n_{0}^{\alpha}$. The Hartree-Fock contributions to $\hbar\Omega
_{00,\text{HF}}$ and $\epsilon_{0,\text{HF}}^{\alpha}$ are linear in the bare
Coulomb potential,~\cite{kochbuch} whereas the correlation (or scattering)
contributions $S_{0}(\omega)$ are given in 2nd Born
approximation.~\cite{khitrova:rmp99} Since we assume a quasi-equilibrium
situation with temporally constant electron and hole populations, we can
analytically evaluate the Fourier transform of the non-Markovian dynamical
equation for the scattering contributions that determine the optical response
to the weak probe pulse. The result for the frequency dependent diagonal part
is
\begin{align}
&  \Lambda_{00}(\omega)=\frac{1}{A^{2}}\sum_{\vec{q}\neq\text{0,}\vec
{k}^{\prime}\text{,}\vec{k}^{\prime\prime}}\sum_{\alpha,\beta=\text{e,h}%
}g(\hbar\omega-\epsilon_{0}^{\bar{\alpha}}-\epsilon_{k^{\prime}}^{\alpha
}+\epsilon_{k^{\prime\prime}+q}^{\beta}-\epsilon_{k^{\prime\prime}}^{\beta
})\nonumber\\
&  \left[  2W_{q}^{2}\,|I_{0k^{\prime}}(q)|^{2}-\delta_{\alpha\beta}%
W_{q}W_{k^{\prime\prime}-k^{\prime}+q}\,I_{0k^{\prime}}(q)I_{0k^{\prime\prime
}}(k^{\prime\prime}-k^{\prime}+q)^{\ast}\right]  \nonumber\\
&  \;\left[  (1-n_{k^{\prime}}^{\alpha})\,n_{k^{\prime\prime}+q}^{\beta
}(1-n_{k^{\prime\prime}}^{\beta})+n_{k^{\prime}}^{\alpha}(1-n_{k^{\prime
\prime}+q}^{\beta})\,n_{k^{\prime\prime}}^{\beta}\right]  \,\ \label{gamma-eh}%
\end{align}
and correspondingly for the off-diagonal parts\textit{\ }$\Lambda
_{0,k^{\prime}},$which can be obtained from Eq.~(\ref{gamma-eh}) by pulling
the $k^{\prime}$ sum out, making the replacements $n_{k^{\prime}}^{\alpha
}\rightarrow n_{0}^{\alpha}$ and using $g(\hbar\omega-\epsilon_{0}^{\alpha
}-\epsilon_{k^{\prime}}^{\bar{\alpha}}+\epsilon_{k^{\prime\prime}+q}^{\beta
}-\epsilon_{k^{\prime\prime}}^{\beta})$. Here, $g(\epsilon)=\lim
_{\eta\rightarrow0}\frac{i}{\epsilon+i\eta}=\pi\delta(\epsilon)+iP\frac
{1}{\epsilon}$ where $\epsilon$ is the energy argument and $P$ denotes the
principal part. Furthermore, writing $\Lambda_{00}=\Gamma_{00}-i\Delta_{00}$,
$\ \Gamma_{00}$\textit{\ }is the dephasing contribution, and $\Delta_{00}$ the
energy renormalization by the correlation contribution that is added to the
Hartree-Fock contribution\textit{.} The screened potential is denoted by
$W_{q}=V_{q}/\varepsilon_{q}$, where $\varepsilon_{q}$ is the Lindhard
dielectric function in static approximation computed using the QW carrier
distributions, and the normalization area is denoted by $A$. The Markovian
limit of Eq.~(\ref{gamma-eh}) is obtained by using $\ g(\epsilon_{0}^{\alpha
}-\epsilon_{k^{\prime}}^{\alpha}+\epsilon_{k^{\prime\prime}+q}^{\beta
}-\epsilon_{k^{\prime\prime}}^{\beta})$, which makes $\Lambda$ frequency
independent. For the QW polarizations $p_{k}$ the equivalent to $\Lambda
_{00}(\omega)$ is $\Lambda_{kk}(\omega)$, which is  computed by using
$g(\hbar\omega-\epsilon_{k}^{\bar{\alpha}}-\epsilon_{k^{\prime}}^{\alpha
}+\epsilon_{k^{\prime\prime}+q}^{\beta}-\epsilon_{k^{\prime\prime}}^{\beta})$
and replacing $I_{0k^{\prime}}(q)\rightarrow I_{kk^{\prime}}(q)=\delta
(k-k^{\prime}+q)$. Thus, the form of Eq.~(\ref{gamma-eh}) describes the
dephasing for both QD and QW states. We
choose a parabolic in-plane confinement potential for the QD and a QW width of
$w=4$ nm, with material parameters typical of the InGaAs/GaAs system. For the
scattering states, we use orthogonalized plane waves for the overlap
integrals. Equations (\ref{p-om})--(\ref{gamma-eh}) couple to the microscopic
QW polarizations $p_{k}$. In deriving the equations of motion for the $p_{k}$,
we have ignored the influence of the QDs on the QW states since this influence
is small for the low QD concentrations considered in the following. The probe
absorption spectra are calculated as in Ref.~\cite{prb:many-body-qds}.

Figure 1 shows absorption spectra obtained by evaluating
Eqs. (\ref{p-om})--(\ref{gamma-eh}) and their counterparts for the QW
polarizations $p_{k} $\ for different carrier densities under
room-temperature ($T=300$K) quasi-equilibrium conditions. Each
spectrum exhibits the QD resonance and the absorption due to the
extended QW states including the QW excitonic resonance.  With
increasing carrier density, the bleaching of the QD resonance is
accompanied by a pronounced red shift and excitation induced
broadening. In contrast, the QW excitonic resonance exhibits only
bleaching and broadening, but no energy shift. Such a red shift of the
QD ground-state resonance has been observed recently in
photoluminescence
experiments.~\cite{matsuda:apl03:carrier-carrier-interaction}

Figure 2 focuses on the line shape of the QD absorption (solid line).
Comparison with a Lorentzian least-squares fit curve (dotted line)
shows that the linewidth of the QD absorption computed including
memory effects is significantly different from a Lorentzian
line. Earlier calculations without memory effects, i.e., essentially
\textit{Markovian} versions of our approach, have also shown
non-Lorentzian excitonic line shapes for bulk and
QWs,~\cite{khitrova:rmp99,kochbuch} but the important point here is
that the absorption linewidth in \textit{non}-Markovian calculation is
smaller than the Markovian result (dashed line). This is surprising
because non-Markovian calculations are known to \textit{broaden} yield
\textit{broader} spectral features than Markovian treatments
because scattering and dephasing are increased compared to the Markovian
treatment by including the memory of the interaction
processes.~\cite{betz:prl2001}

Figure 3 explains why the \textit{non}-Markovian calculation can lead
to a smaller dephasing than the Markovian calculation. It shows the
real part of Eq.~(\ref{gamma-eh}), which determines the dephasing,
i.e., the broadening of the absorption spectrum. The $\omega$
dependence of $\Gamma=\mathrm{Re}\Lambda_{00}$ shows that the
non-Markovian calculation does not yield energy conservation among the
single-particle states connected by the scattering process, but
introduces an additional $\hbar\omega$ contribution. The non-Markovian
calculation can therefore lead to a larger dephasing, but in the
present case this is only true for $\hbar\omega\geq\epsilon_{0}$
because in this region more QW states can satisfy the argument of the
$\delta$ function, which occurs in the real part of
Eq.~(\ref{gamma-eh}). For energies smaller than the unexcited dot
resonance $\hbar\omega\leq\epsilon_{0}$, the opposite effect occurs
because less QW energies in Eq.~(\ref{gamma-eh}) satisfy the argument
of the $\delta$ function. Thus, the QD dephasing is determined the
value of $\Gamma(\omega)$ \textit{at the QD resonance}. As can be seen
in Fig.~1 the QD resonance undergoes a redshift and
\textquotedblleft picks\textquotedblright\ the dephasing at the
energies marked by dots in Fig.~2. The excitation induced shift of the
QD resonance is the reason why the non-Markovian calculation leads to a
narrower line than the Markovian calculation.

Figure 4 summarizes the results for the excitation dependence of the
dephasing. Plotted is the full width at half maximum (FWHM) versus
carrier density. The main figure shows the difference between the full
(solid curve) and Markovian (dashed curve) calculations. While both
curves exhibit a linear dependence of spectral width on carrier
density, using the Markovian approximation overestimates both the
magnitude and the carrier density dependence. As explained above, this
result is a consequence of the frequency dependent effective dephasing
rate and the excitation induced red shift of the dot resonance. The
inset compares these results to the QW broadening. In addition to an
appreciably larger broadening because of a larger phase space that
contributes to the scattering integral, the QW spectral width has a
superlinear dependence on carrier density, because the QW exciton
experiences the excitation induced dephasing at the same energy
without counteracting this trend by undergoing a red shift like the QD
resonance. Comparison with experiment is best done in terms of the
change in spectral width with carrier density, as it eliminates the
background due to carrier-phonon scattering. We extracted values of
$3$ to $6\times 10^{-18}\operatorname{meV}/$cm$^{-3}$ from single QD
measurements,~\cite{matsuda:apl03:carrier-carrier-interaction} where
the variation arises from dot-to-dot fluctuations and uncertainty in
carrier density. This range of values compares favorably with our
prediction of $2\times10^{-18}$ meV$/$cm$^{-3}$ (slope of solid curve
in Fig. 4), where we used the QW width of 4 nm and assumed
$\epsilon_{R}=4$ meV.

In conclusion, we present a non-Markovian quantum-kinetic analysis of
the excitation dependent energy renormalization and broadening of QD
states that are electronically coupled to a continuum of states, which
we model as a QW.  The dot resonance displays a density-dependent red
shift, which is the counterpart of the band-gap shift for the QW
states, whereas the QW excitonic resonance is stationary due to the
compensation of gap shift and reduction of the exciton binding
energy. The non-Markovian calculation introduces a frequency dependent
damping, and its value at the excitation dependent QD resonance
determines the magnitude of the polarization dephasing, i.e., the
linewidth of the QD resonance. The QD resonance shows a pronounced
excitation induced dephasing, but the excitation induced shift of the
resonance counteracts the overall increase of the damping to some
extent. Compared to the QD resonance, the QW exciton's damping is
larger due to a larger phase space available for the scattering
processes, and its stationary resonance experiences the excitation
induced increase of the damping at the QW excitonic resonance and thus
a larger excitation induced dephasing.

\begin{acknowledgments}
We thank W. Hoyer, M. Kira, and F. Jahnke for helpful discussions.
This work was funded in part by the United States Department of Energy under
contract DE-AC04-94AL85000. WWC was supported in part by the Senior Scientist
Program of the Humboldt Foundation. SWK thanks the Deutsche
Forschungsgemeinschaft, the Humboldt Foundation and the Max-Planck Society for support.
\end{acknowledgments}

\bibliographystyle{apsrev}
\bibliography{../../../citation/dephasing,../../../citation/lwi,../../../citation/hcs_pubs,../../../citation/dotbib}

\begin{thebibliography}{21}
\expandafter\ifx\csname natexlab\endcsname\relax\def\natexlab#1{#1}\fi
\expandafter\ifx\csname bibnamefont\endcsname\relax
  \def\bibnamefont#1{#1}\fi
\expandafter\ifx\csname bibfnamefont\endcsname\relax
  \def\bibfnamefont#1{#1}\fi
\expandafter\ifx\csname citenamefont\endcsname\relax
  \def\citenamefont#1{#1}\fi
\expandafter\ifx\csname url\endcsname\relax
  \def\url#1{\texttt{#1}}\fi
\expandafter\ifx\csname urlprefix\endcsname\relax\def\urlprefix{URL }\fi
\providecommand{\bibinfo}[2]{#2}
\providecommand{\eprint}[2][]{\url{#2}}

\bibitem[{\citenamefont{Michler et~al.}(2000)\citenamefont{Michler, Imamoglu,
  Mason, Carson, Strouse, and Buratto}}]{michler:nature-quant-correlation}
\bibinfo{author}{\bibfnamefont{P.}~\bibnamefont{Michler}},
  \bibinfo{author}{\bibfnamefont{A.}~\bibnamefont{Imamoglu}},
  \bibinfo{author}{\bibfnamefont{M.~D.} \bibnamefont{Mason}},
  \bibinfo{author}{\bibfnamefont{P.~J.} \bibnamefont{Carson}},
  \bibinfo{author}{\bibfnamefont{G.~F.} \bibnamefont{Strouse}},
  \bibnamefont{and} \bibinfo{author}{\bibfnamefont{S.~K.}
  \bibnamefont{Buratto}}, \bibinfo{journal}{Nature}
  \textbf{\bibinfo{volume}{406}}, \bibinfo{pages}{968} (\bibinfo{year}{2000}).

\bibitem[{\citenamefont{Bayer et~al.}(2001)\citenamefont{Bayer, Reinecke,
  Weidner, Larionov, McDonald, and Forchel}}]{bayer:prl01:spont-em}
\bibinfo{author}{\bibfnamefont{M.}~\bibnamefont{Bayer}},
  \bibinfo{author}{\bibfnamefont{T.~L.} \bibnamefont{Reinecke}},
  \bibinfo{author}{\bibfnamefont{F.}~\bibnamefont{Weidner}},
  \bibinfo{author}{\bibfnamefont{A.}~\bibnamefont{Larionov}},
  \bibinfo{author}{\bibfnamefont{A.}~\bibnamefont{McDonald}}, \bibnamefont{and}
  \bibinfo{author}{\bibfnamefont{A.}~\bibnamefont{Forchel}},
  \bibinfo{journal}{Phys.\ Rev.\ Lett.} \textbf{\bibinfo{volume}{86}},
  \bibinfo{pages}{3168} (\bibinfo{year}{2001}).

\bibitem[{\citenamefont{Li et~al.}(2003)\citenamefont{Li, Wu, Steel, Gammon,
  Stievater, Katzer, Park, Piermarocchi, and Sham}}]{steel:all-optical-gate}
\bibinfo{author}{\bibfnamefont{X.~Q.} \bibnamefont{Li}},
  \bibinfo{author}{\bibfnamefont{Y.~W.} \bibnamefont{Wu}},
  \bibinfo{author}{\bibfnamefont{D.}~\bibnamefont{Steel}},
  \bibinfo{author}{\bibfnamefont{D.}~\bibnamefont{Gammon}},
  \bibinfo{author}{\bibfnamefont{T.~H.} \bibnamefont{Stievater}},
  \bibinfo{author}{\bibfnamefont{D.~S.} \bibnamefont{Katzer}},
  \bibinfo{author}{\bibfnamefont{D.}~\bibnamefont{Park}},
  \bibinfo{author}{\bibfnamefont{C.}~\bibnamefont{Piermarocchi}},
  \bibnamefont{and} \bibinfo{author}{\bibfnamefont{L.~J.} \bibnamefont{Sham}},
  \bibinfo{journal}{Science} \textbf{\bibinfo{volume}{301}},
  \bibinfo{pages}{809} (\bibinfo{year}{2003}).

\bibitem[{\citenamefont{Sugawara}(1999)}]{Sugawara-book}
\bibinfo{author}{\bibfnamefont{M.}~\bibnamefont{Sugawara}},
  \emph{\bibinfo{title}{Self-Assembled {InGaAs/GaAs} Quantum Dots}}
  (\bibinfo{publisher}{Academic}, \bibinfo{address}{San Diego},
  \bibinfo{year}{1999}).

\bibitem[{\citenamefont{Bimberg et~al.}(1999)\citenamefont{Bimberg, Grundmann,
  and Ledentsov}}]{Bimberg-book}
\bibinfo{author}{\bibfnamefont{D.}~\bibnamefont{Bimberg}},
  \bibinfo{author}{\bibfnamefont{M.}~\bibnamefont{Grundmann}},
  \bibnamefont{and} \bibinfo{author}{\bibfnamefont{N.~N.}
  \bibnamefont{Ledentsov}}, \emph{\bibinfo{title}{Quantum Dot
  Heterostructures}} (\bibinfo{publisher}{Wiley}, \bibinfo{address}{New York},
  \bibinfo{year}{1999}).

\bibitem[{\citenamefont{Uskov et~al.}(2000)\citenamefont{Uskov, Jauho,
  Tromborg, Mork, and Lang}}]{uskov-jauho:dotdephasing-phonon}
\bibinfo{author}{\bibfnamefont{A.~V.} \bibnamefont{Uskov}},
  \bibinfo{author}{\bibfnamefont{A.~P.} \bibnamefont{Jauho}},
  \bibinfo{author}{\bibfnamefont{B.}~\bibnamefont{Tromborg}},
  \bibinfo{author}{\bibfnamefont{J.}~\bibnamefont{Mork}}, \bibnamefont{and}
  \bibinfo{author}{\bibfnamefont{R.}~\bibnamefont{Lang}},
  \bibinfo{journal}{Phys.\ Rev.\ Lett.} \textbf{\bibinfo{volume}{85}},
  \bibinfo{pages}{1516} (\bibinfo{year}{2000}).

\bibitem[{\citenamefont{Schneider et~al.}(2002)\citenamefont{Schneider, Chow,
  and Koch}}]{prb02:anomalous-dispersion}
\bibinfo{author}{\bibfnamefont{H.~C.} \bibnamefont{Schneider}},
  \bibinfo{author}{\bibfnamefont{W.~W.} \bibnamefont{Chow}}, \bibnamefont{and}
  \bibinfo{author}{\bibfnamefont{S.~W.} \bibnamefont{Koch}},
  \bibinfo{journal}{Phys.\ Rev.~B} \textbf{\bibinfo{volume}{66}},
  \bibinfo{pages}{041310} (\bibinfo{year}{2002}).

\bibitem[{\citenamefont{Sugawara et~al.}(2000)\citenamefont{Sugawara, Mukai,
  Nakata, Ishikawa, and Sakamoto}}]{sugawara:homogeneous_broadening}
\bibinfo{author}{\bibfnamefont{M.}~\bibnamefont{Sugawara}},
  \bibinfo{author}{\bibfnamefont{K.}~\bibnamefont{Mukai}},
  \bibinfo{author}{\bibfnamefont{Y.}~\bibnamefont{Nakata}},
  \bibinfo{author}{\bibfnamefont{H.}~\bibnamefont{Ishikawa}}, \bibnamefont{and}
  \bibinfo{author}{\bibfnamefont{A.}~\bibnamefont{Sakamoto}},
  \bibinfo{journal}{Phys.\ Rev.~B} \textbf{\bibinfo{volume}{61}},
  \bibinfo{pages}{7595} (\bibinfo{year}{2000}).

\bibitem[{\citenamefont{Borri et~al.}(2001)\citenamefont{Borri, Langbein,
  Schneider, Woggon, Sellin, Ouyang, and Bimberg}}]{borri:dotdephasing-prl}
\bibinfo{author}{\bibfnamefont{P.}~\bibnamefont{Borri}},
  \bibinfo{author}{\bibfnamefont{W.}~\bibnamefont{Langbein}},
  \bibinfo{author}{\bibfnamefont{S.}~\bibnamefont{Schneider}},
  \bibinfo{author}{\bibfnamefont{U.}~\bibnamefont{Woggon}},
  \bibinfo{author}{\bibfnamefont{R.~L.} \bibnamefont{Sellin}},
  \bibinfo{author}{\bibfnamefont{D.}~\bibnamefont{Ouyang}}, \bibnamefont{and}
  \bibinfo{author}{\bibfnamefont{D.}~\bibnamefont{Bimberg}},
  \bibinfo{journal}{Phys.\ Rev.\ Lett.} \textbf{\bibinfo{volume}{87}},
  \bibinfo{pages}{157401} (\bibinfo{year}{2001}).

\bibitem[{\citenamefont{Guenther et~al.}(2002)\citenamefont{Guenther, Lienau,
  Elsaesser, Glanemann, Axt, Kuhn, Eshlaghi, and
  Wieck}}]{guenther:quantum-dot-response}
\bibinfo{author}{\bibfnamefont{T.}~\bibnamefont{Guenther}},
  \bibinfo{author}{\bibfnamefont{C.}~\bibnamefont{Lienau}},
  \bibinfo{author}{\bibfnamefont{T.}~\bibnamefont{Elsaesser}},
  \bibinfo{author}{\bibfnamefont{M.}~\bibnamefont{Glanemann}},
  \bibinfo{author}{\bibfnamefont{V.~M.} \bibnamefont{Axt}},
  \bibinfo{author}{\bibfnamefont{T.}~\bibnamefont{Kuhn}},
  \bibinfo{author}{\bibfnamefont{S.}~\bibnamefont{Eshlaghi}}, \bibnamefont{and}
  \bibinfo{author}{\bibfnamefont{A.~D.} \bibnamefont{Wieck}},
  \bibinfo{journal}{Phys.\ Rev.\ Lett.} \textbf{\bibinfo{volume}{89}},
  \bibinfo{pages}{057401} (\bibinfo{year}{2002}).

\bibitem[{\citenamefont{Toda et~al.}(1999)\citenamefont{Toda, Moriwaki,
  Nishioka, and Arakawa}}]{toda:carrier-relaxation}
\bibinfo{author}{\bibfnamefont{Y.}~\bibnamefont{Toda}},
  \bibinfo{author}{\bibfnamefont{O.}~\bibnamefont{Moriwaki}},
  \bibinfo{author}{\bibfnamefont{M.}~\bibnamefont{Nishioka}}, \bibnamefont{and}
  \bibinfo{author}{\bibfnamefont{Y.}~\bibnamefont{Arakawa}},
  \bibinfo{journal}{Phys.\ Rev.\ Lett.} \textbf{\bibinfo{volume}{82}},
  \bibinfo{pages}{4114} (\bibinfo{year}{1999}).

\bibitem[{\citenamefont{Urayama et~al.}(2001)\citenamefont{Urayama, Norris,
  Singh, and Bhattacharya}}]{urayama:prl01:phonon-bottleneck}
\bibinfo{author}{\bibfnamefont{J.}~\bibnamefont{Urayama}},
  \bibinfo{author}{\bibfnamefont{T.~B.} \bibnamefont{Norris}},
  \bibinfo{author}{\bibfnamefont{J.}~\bibnamefont{Singh}}, \bibnamefont{and}
  \bibinfo{author}{\bibfnamefont{P.}~\bibnamefont{Bhattacharya}},
  \bibinfo{journal}{Phys.\ Rev.\ Lett.} \textbf{\bibinfo{volume}{86}},
  \bibinfo{pages}{4930} (\bibinfo{year}{2001}).

\bibitem[{\citenamefont{Banyai et~al.}(1995)\citenamefont{Banyai, Tran~Thoai,
  Reitsamer, Haug, Steinbach, Wehner, Wegener, Marschner, and
  Stolz}}]{banyai:exciton-lophonon}
\bibinfo{author}{\bibfnamefont{L.}~\bibnamefont{Banyai}},
  \bibinfo{author}{\bibfnamefont{D.~B.} \bibnamefont{Tran~Thoai}},
  \bibinfo{author}{\bibfnamefont{E.}~\bibnamefont{Reitsamer}},
  \bibinfo{author}{\bibfnamefont{H.}~\bibnamefont{Haug}},
  \bibinfo{author}{\bibfnamefont{D.}~\bibnamefont{Steinbach}},
  \bibinfo{author}{\bibfnamefont{M.~U.} \bibnamefont{Wehner}},
  \bibinfo{author}{\bibfnamefont{M.}~\bibnamefont{Wegener}},
  \bibinfo{author}{\bibfnamefont{T.}~\bibnamefont{Marschner}},
  \bibnamefont{and} \bibinfo{author}{\bibfnamefont{W.}~\bibnamefont{Stolz}},
  \bibinfo{journal}{Phys.\ Rev.\ Lett.} \textbf{\bibinfo{volume}{75}},
  \bibinfo{pages}{2188} (\bibinfo{year}{1995}).

\bibitem[{\citenamefont{Haug and Koch}(2004)}]{kochbuch}
\bibinfo{author}{\bibfnamefont{H.}~\bibnamefont{Haug}} \bibnamefont{and}
  \bibinfo{author}{\bibfnamefont{S.~W.} \bibnamefont{Koch}},
  \emph{\bibinfo{title}{Quantum Theory of the Optical and Electronic Properties
  of Semiconductors}} (\bibinfo{publisher}{World Scientific},
  \bibinfo{address}{Singapore}, \bibinfo{year}{2004}).

\bibitem[{\citenamefont{Schneider et~al.}(2001)\citenamefont{Schneider, Chow,
  and Koch}}]{prb:many-body-qds}
\bibinfo{author}{\bibfnamefont{H.~C.} \bibnamefont{Schneider}},
  \bibinfo{author}{\bibfnamefont{W.~W.} \bibnamefont{Chow}}, \bibnamefont{and}
  \bibinfo{author}{\bibfnamefont{S.~W.} \bibnamefont{Koch}},
  \bibinfo{journal}{Phys.\ Rev.~B} \textbf{\bibinfo{volume}{64}},
  \bibinfo{pages}{115315} (\bibinfo{year}{2001}).

\bibitem[{\citenamefont{Sch\"afer and Wegener}(2002)}]{schaefer-book}
\bibinfo{author}{\bibfnamefont{W.}~\bibnamefont{Sch\"afer}} \bibnamefont{and}
  \bibinfo{author}{\bibfnamefont{M.}~\bibnamefont{Wegener}},
  \emph{\bibinfo{title}{Semiconductor Optics and Transport Phenomena}}
  (\bibinfo{publisher}{Springer}, \bibinfo{year}{2002}).

\bibitem[{\citenamefont{Haug and Jauho}(1996)}]{haug-jauho}
\bibinfo{author}{\bibfnamefont{H.}~\bibnamefont{Haug}} \bibnamefont{and}
  \bibinfo{author}{\bibfnamefont{A.-P.} \bibnamefont{Jauho}},
  \emph{\bibinfo{title}{Quantum Kinetics in Transport and Optics of
  Semiconductors}}, vol. \bibinfo{volume}{123} of
  \emph{\bibinfo{series}{Springer Series in Solid-State Sciences}}
  (\bibinfo{publisher}{Springer}, \bibinfo{address}{Berlin},
  \bibinfo{year}{1996}).

\bibitem[{\citenamefont{Binder and Koch}(1995)}]{binder-koch}
\bibinfo{author}{\bibfnamefont{R.}~\bibnamefont{Binder}} \bibnamefont{and}
  \bibinfo{author}{\bibfnamefont{S.~W.} \bibnamefont{Koch}},
  \bibinfo{journal}{Progress Quant. Electronics} \textbf{\bibinfo{volume}{19}},
  \bibinfo{pages}{307} (\bibinfo{year}{1995}).

\bibitem[{\citenamefont{Khitrova et~al.}(1999)\citenamefont{Khitrova, Gibbs,
  Jahnke, Kira, and Koch}}]{khitrova:rmp99}
\bibinfo{author}{\bibfnamefont{G.}~\bibnamefont{Khitrova}},
  \bibinfo{author}{\bibfnamefont{H.~M.} \bibnamefont{Gibbs}},
  \bibinfo{author}{\bibfnamefont{F.}~\bibnamefont{Jahnke}},
  \bibinfo{author}{\bibfnamefont{M.}~\bibnamefont{Kira}}, \bibnamefont{and}
  \bibinfo{author}{\bibfnamefont{S.~W.} \bibnamefont{Koch}},
  \bibinfo{journal}{Rev.\ Mod.\ Phys.} \textbf{\bibinfo{volume}{71}},
  \bibinfo{pages}{1591} (\bibinfo{year}{1999}).

\bibitem[{\citenamefont{Matsuda et~al.}(2003)\citenamefont{Matsuda, Ikeda,
  Saiki, Saito, and Nishi}}]{matsuda:apl03:carrier-carrier-interaction}
\bibinfo{author}{\bibfnamefont{K.}~\bibnamefont{Matsuda}},
  \bibinfo{author}{\bibfnamefont{K.}~\bibnamefont{Ikeda}},
  \bibinfo{author}{\bibfnamefont{T.}~\bibnamefont{Saiki}},
  \bibinfo{author}{\bibfnamefont{H.}~\bibnamefont{Saito}}, \bibnamefont{and}
  \bibinfo{author}{\bibfnamefont{K.}~\bibnamefont{Nishi}},
  \bibinfo{journal}{Appl.\ Phys.\ Lett.} \textbf{\bibinfo{volume}{83}},
  \bibinfo{pages}{2250} (\bibinfo{year}{2003}).

\bibitem[{\citenamefont{Betz et~al.}(2001)\citenamefont{Betz, Goger, Laubereau,
  Gartner, Banyai, Haug, Ortner, Becker, and Leitenstorfer}}]{betz:prl2001}
\bibinfo{author}{\bibfnamefont{M.}~\bibnamefont{Betz}},
  \bibinfo{author}{\bibfnamefont{G.}~\bibnamefont{Goger}},
  \bibinfo{author}{\bibfnamefont{A.}~\bibnamefont{Laubereau}},
  \bibinfo{author}{\bibfnamefont{P.}~\bibnamefont{Gartner}},
  \bibinfo{author}{\bibfnamefont{L.}~\bibnamefont{Banyai}},
  \bibinfo{author}{\bibfnamefont{H.}~\bibnamefont{Haug}},
  \bibinfo{author}{\bibfnamefont{K.}~\bibnamefont{Ortner}},
  \bibinfo{author}{\bibfnamefont{C.~R.} \bibnamefont{Becker}},
  \bibnamefont{and}
  \bibinfo{author}{\bibfnamefont{A.}~\bibnamefont{Leitenstorfer}},
  \bibinfo{journal}{Phys.\ Rev.\ Lett.} \textbf{\bibinfo{volume}{86}},
  \bibinfo{pages}{4684} (\bibinfo{year}{2001}).

\end{thebibliography}

\newpage

\begin{figure}[t]
\centering \resizebox{0.5\columnwidth}{!}{\includegraphics{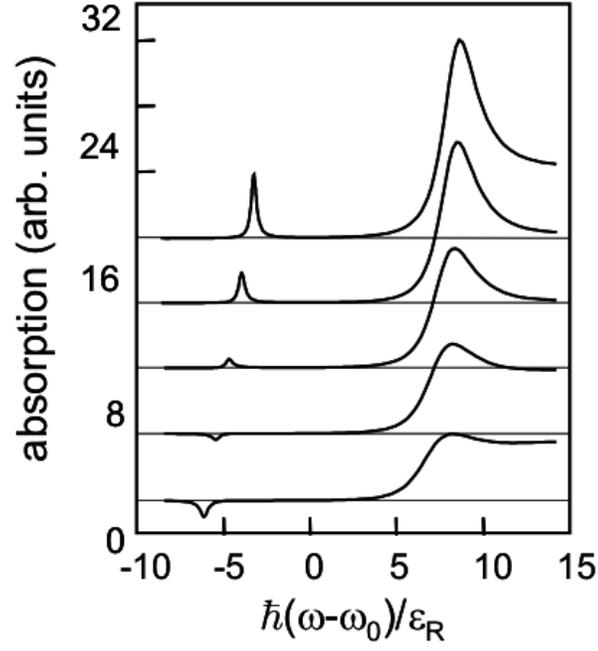}}
\caption{Room-temperature absorption spectra for the combined
quantum-dot/quantum-well system for carrier densities 2 to $4\times
10^{11}\,\text{cm}^{-2}$in intervals of $5\times10^{10}\,\text{cm}^{-2}$ (top
to bottom, the base line of the spectra is shifted). The energy unit is the
3-dimensional exciton binding energy $\epsilon_{R}$, and the zero is at the
bare quantum-dot excitonic transition $\hbar\omega_{0}$.}%
\end{figure}
\begin{figure}[b]
\centering \resizebox{0.5\columnwidth}{!}{\includegraphics{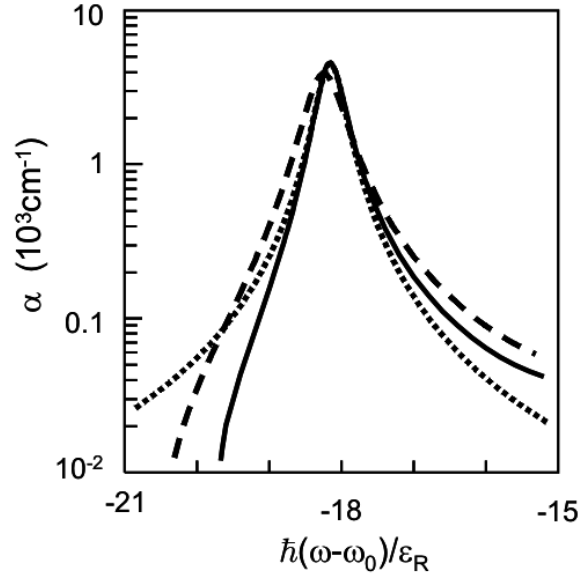}}
\caption{Absorption
spectrum around quantum-dot resonance for carrier density $2.5\times
10^{11}\,\text{cm}^{-2}$ calculated using the full non-Markovian scattering
contributions (solid curve), and the Markov approximation (dashed curve). The
dotted curve is obtained from a least-squares fit with a Lorentzian.}%
\end{figure}

\begin{figure}[t]
\centering \resizebox{0.5\columnwidth}{!}{\includegraphics{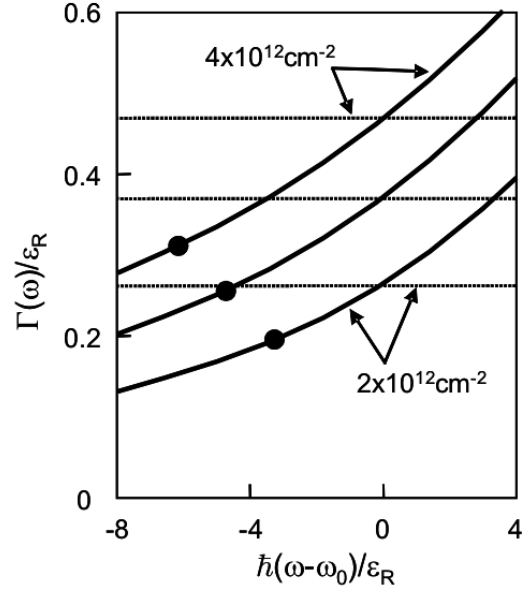}}
\caption{$\Gamma(\omega)= \mathrm{Re}\,\Lambda_{00}(\omega)$
[Eq.~(\protect\ref{gamma-eh})] for three different carrier
densities. The respective renormalized quantum-dot resonance is marked
by a dot.}
\end{figure}

\begin{figure}[b]
\centering \resizebox{0.5\columnwidth}{!}{\includegraphics{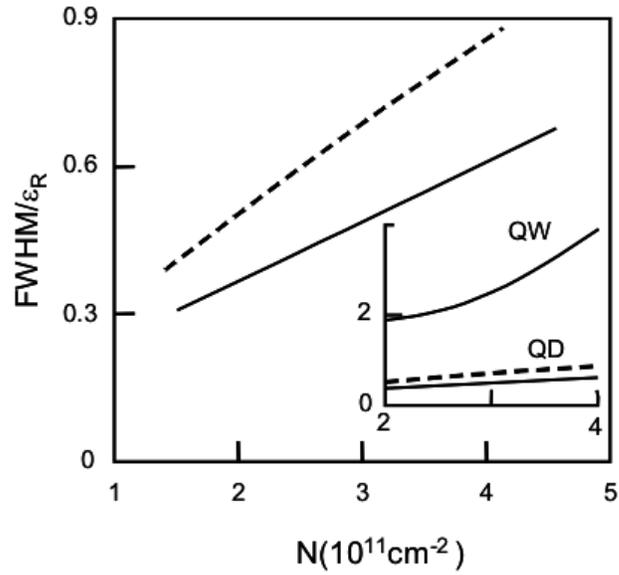}}
\caption{Excitation dependent broadening of the quantum-dot and quantum-well
resonance lines calculated using the full non-Markovian scattering
contributions (solid lines) and the Markov approximation (dashed line).
Inset:\ Comparison to excitation induced broadening of the quantum-well
excitonic resonance. The energy unit is the 3-d exciton binding energy.}%
\end{figure}

\end{document}